   \documentclass[final,5p,times,twocolumn,authoryear]{elsarticle}

\usepackage{amssymb}
\journal{Applied Radiation and Isotopes}

\begin{document}

%%-----------------------------------------------------------------------------
\begin{frontmatter}

%% Title, authors and addresses

%% use the tnoteref command within \title for footnotes;
%% use the tnotetext command for theassociated footnote;
%% use the fnref command within \author or \address for footnotes;
%% use the fntext command for theassociated footnote;
%% use the corref command within \author for corresponding author footnotes;
%% use the cortext command for theassociated footnote;
%% use the ead command for the email address,
%% and the form \ead[url] for the home page:
%% \title{Title\tnoteref{label1}}
%% \tnotetext[label1]{}
%% \author{Name\corref{cor1}\fnref{label2}}
%% \ead{email address}
%% \ead[url]{home page}
%% \fntext[label2]{}
%% \cortext[cor1]{}
%% \address{Address\fnref{label3}}
%% \fntext[label3]{}

 \title{ A novel approach to medical radioisotope production using inverse kinematics: \\
       a successful production test of the theranostic radionuclide $^{67}$Cu }

%% \title{ A novel production approach of medical radioisotopes: %%% exemplified in the
%%        a successful production test of the theranostic radionuclide $^{67}$Cu in inverse kinematics}

%% use optional labels to link authors explicitly to addresses:
%% \author[label1,label2]{}
%% \address[label1]{}
%% \address[label2]{}

\author{G.A. Souliotis$^{a,1}$, M.R.D. Rodrigues$^{b,e}$, K. Wang$^{b}$, V. Iacob$^{b}$, N. Nica$^{b}$, B. Roeder$^{b}$, 
        G. Tabacaru$^{b}$, 
        \\ M. Yu$^{c}$, P. Zanotti-Fregonara$^{c}$, A. Bonasera$^{b,d}$  }

\fntext[myfootnote]{ Corresponding author: soulioti@chem.uoa.gr}

\address{ $^{a}$ Laboratory of Physical Chemistry, Department of Chemistry,
                 National and Kapodistrian University of Athens, Athens 15771, Greece }

\address{ $^{b}$ Cyclotron Institute, Texas A \& M University, College Station, Texas 77048, USA}

\address{ $^{c}$ Houston Methodist Research Institute, Houston, Texas , 77030, USA }

\address{ $^{d}$ Laboratori Nazionali del Sud, INFN, Catania 95123, Italy }

\address{ $^{e}$ Instituto de F\'isica, Universidade de S\~ao Paulo, S\~ao Paulo 05508-090, Brazil }

%%-------------------------------------------------------------------------
\begin{abstract}

A novel method for the production of important medical radioisotopes has been developed.
The approach is based on performing the nuclear reaction in inverse kinematics,
namely sending a heavy-ion beam of appropriate energy on a light target (e.g. H, d, He) and 
collecting the isotope of interest.
%%%
In this work, as a proof-of-concept, we studied the production of the theranostic
radionuclide $^{67}$Cu (T$_{1/2}$=62 h) via the reaction of a $^{70}$Zn beam at 15 MeV/nucleon with 
a hydrogen gas target. The $^{67}$Cu radionuclide, alongside other coproduced isotopes, was collected 
after the gas target on an Al catcher foil and their radioactivity was measured by off-line 
$\gamma$-ray analysis. 
%%%
After 36 h from the end of the irradiation, apart from the product of interest $^{67}$Cu,
the main radioimpurity coming from the $^{70}$Zn+p reaction was $^{69m}$Zn (T$_{1/2}$=13.8 h) 
that can be reduced by further radio-cooling. 
%% Other identified radioimpurities are understood to come from the interaction of the beam with 
%% the window material and the catcher and can be eliminated by careful tuning of the parameters 
%% of the setup.
%%Our study suggests the possibility of producing important non-standard radionuclides %%% like $^{67}$Cu
%%with high radionuclide purity with the approach of inverse kinematics.
%%%
Moreover, along with the radionuclide of interest produced in inverse kinematics, the production of  additional
radioisotopes is possible by making use of the forward-focused neutrons from the reaction and letting them 
interact with a secondary target.  A preliminary successful test of this concept was realized in the present study. 
%%%
The main requirement to obtain activities appropriate for preclinical studies is the development
of high-intensity heavy-ion primary beams.

%%%

\end{abstract}
%%%------------------------------------------------------------------------
\begin{keyword}
%% keywords here, in the form: keyword \sep keyword
%%
%% PACS codes here, in the form: \PACS code \sep code
%%
%% MSC codes here, in the form: \MSC code \sep code
%% or \MSC[2008] code \sep code (2000 is the default)

\end{keyword}

\end{frontmatter}

%%----------------------------------------------------------------------------
%% \linenumbers

%%
%% main text

\section{Introduction}
\label{intro}

Medical radionuclides play a central role in nuclear medicine in the fields
of diagnostic imaging and radioimmunotherapy (RIT) (\cite{Qaim-2017,Srivastava-2014,Stocklin-1995}).
%%% internal therapy (endotherapy).
%%%
Radionuclides emitting low-range highly ionizing radiation 
( $\beta^{-}$ or $\alpha$ particles, Auger or conversion electrons)  are 
essential for RIT approaches.
%%%
Apart from a number of standard radionuclides, %%%%%% (\cite{Qaim-2017}),
currently the  $\beta^{-}$ emitters 
$^{47}$Sc (T$_{1/2}$=3.4 d),
$^{67}$Cu (T$_{1/2}$=2.6 d),
$^{105}$Rh (T$_{1/2}$=1.5 d),
$^{161}$Tb (T$_{1/2}$=6.9 d) and
$^{186}$Re (T$_{1/2}$=3.7 d) (\cite{Champion-2016,Qaim-2017})
are increasingly interesting.

Specifically,  $^{67}$Cu, the longest-lived radioisotope of copper,  is ideally 
suited for  both radioimmunotherapy and imaging for several reasons (\cite{Asabella-2014}).
%%%
First, from a chemical perspective, copper is an essential trace element for most organisms 
and specifically for humans as it takes part in important biochemical processes (\cite{Linder-1991}).
%%%
The coordination chemistry of Cu has been well established (\cite{Price-2014}).
Copper can be linked to antibodies, proteins and other biologically important molecules 
(\cite{Follacchio-2018,Ting-2009,Schubiger-1996,Sugo-2017}).
%%%
The nuclide $^{67}$Cu can be combined with the same type of radiopharmaceuticals as 
$^{64}$Cu (T$_{1/2}$=12.7 h)  or $^{61}$Cu (T$_{1/2}$=3.3 h) leading
to efficient theranostic pairs (\cite{Zimmermann-2003}).

The half-life of $^{67}$Cu (62 h) is appropriate to deliver a high dose rate  to the tumor.
Furthermore, its $\beta^{-}$ decay (E$_{e,ave}$ = 141 keV) is followed by the emission of
soft $\gamma$ radiation of 185 keV (48.7\%), 93 keV (16\%) and 91 keV (7\%). This 
makes $^{67}$Cu suitable for imaging the radiotracer distribution by single-photon emission 
computed tomography (SPECT) using the cameras widely developed for the 140 keV 
$\gamma$ rays of  $^{99m}$Tc.
%%%
Compared to the standard RIT radioisotope $^{90}$Y (T$_{1/2}$=64 h, E$_{e,max}$=2.28 MeV),
which is a pure $\beta^{-}$ emitter, $^{67}$Cu offers the possibility of SPECT imaging
and treatment of smaller size tumors (up to 4 mm, compared to 12 mm in the case of $^{90}$Y).
%%%
In addition, $^{67}$Cu compares favorably with another standard radioisotope,
$^{131}$I (T$_{1/2}$=8.0 d, E$_{e,max}$=0.61 MeV),
which has a longer half-life and emits higher energy $\gamma$ rays (0.364 MeV, 82\%) 
and thus, may increase the undesired dose to the patient and the medical personnel.

It is noteworthy that while the other radioisotopes of copper, especially $^{64}$Cu, have already 
been used in radiopharmaceuticals for a wide range of preclinical and clinical studies
(\cite{Follacchio-2018,Peng-2006}), $^{67}$Cu has been used  in a rather limited number
of studies, albeit with very promising results %%% due to its favorable  decay properties
(\cite{Jin-2017,Katz-1990,Knogler-2007,Novak-2002}).
%%%
The main factor limiting the wider preclinical and clinical use is its 
limited availability (\cite{Smith-2012}).

%%%----------------------------------------------------------------------------------------

The  production of $^{67}$Cu in nuclear reactors started about 50 years ago
(\cite{OBrien-1969}) and continues until present in several reactor  facilities
(e.g. \cite{Johnsen-2015,Uddin-2014,Mirzadeh-1986}).
%%%
Recently, however, the main focus has shifted to methods based on particle accelerators
(\cite{Smith-2012}).
%%%
Presently, the main production route is via the reaction $^{68}$Zn(p,2p)$^{67}$Cu
(\cite{Katabuchi-2008,Medvedev-2008,Pupillo-2018,Stoll-2002}).
This approach is   based on the use of intense medium-energy (E$_p$=70--100 MeV) proton beams
that are available by several particle accelerators, including medium-energy cyclotrons. 
However, these multipurpose facilities cannot dedicate all their beam time to radioisotope
production.

Other production routes based on lower-energy charged particle reactions are 
%%%
$^{70}$Zn(p,$\alpha$)$^{67}$Cu  (\cite{Hilgers-2003,Jamriska-1995,Kastleiner-1999}), 
%%%
$^{70}$Zn(d,$\alpha$n)$^{67}$Cu (\cite{Kozempel-2012}) and
%%%
$^{nat}$Zn(d,x)$^{67}$Cu (\cite{Hosseini-2017}), 
%%%
$^{64}$Ni($\alpha$,p)$^{67}$Cu (\cite{Ohya-2018,Skakun-2004}).
%%%
Moreover, production routes based on reactions induced by accelerator-produced neutrons 
have been applied (\cite{Kawabata-2015,Kin-2013,Sato-2014,Spahn-2004}).
In addition, $^{67}$Cu has been produced in photonuclear reactions using
bremsstrahlung photons from high-intensity electron linacs
(\cite{Gopalakrishna-2018,Starovoitova-2014,Starovoitova-2015,Yagi-1978}).

%%%%------------------------------------------------------------------------------------------
Finally,  isotope harvesting in projectile fragmentation facilities has been suggested as
an alternative source of medical isotopes. A proof-of-concept was presented 
in the recent work by Mastren et al. (\cite{Mastren-2014,Mastren-2015}) at the NSCL. 
This work  involved  harvesting  and 
separation of $^{67}$Cu from a mixture of projectile fragments stopped in an aqueous beam-collection
system. The $^{67}$Cu radionuclide separation was followed by radiolabelling
and biodistribution studies.
 
A general characteristic of all the previous production methods is the fact that
the desired radioisotope of $^{67}$Cu is produced inside the target material
which can be moderately or highly expensive, depending on the setup and approach. 
In regard to the production of $^{67}$Cu, the natural abundance of $^{68}$Zn is 18.45\%, 
$^{70}$Zn is 0.61\%  and $^{64}$Ni is 0.93\%. 
%%%
Thus, in these cases, an efficient analysis scheme is necessary for the collection of
the desired $^{67}$Cu isotope and the recovery of the target material for subsequent re-use in
the production scheme (\cite{Smith-2012}).

Along with this traditional scheme, in the isotope-harvesting route,  
the isotope of interest has to be separated from a very broad range of radioisotopes 
of other elements that are abundantly coproduced in a projectile fragmentation reaction.
Thus, an appropriate multistep separation process is necessary (\cite{Mastren-2015}).

%%%------------------------------------------------------------------------------------------------
%%% In this paper:

In this paper, we present an innovative approach for the production of 
medical radioisotopes based on inverse-kinematics nuclear reactions,
that is, sending a heavy-ion beam on a light target and  collecting the 
radioisotope after the target.
%%%
The main advantage of using an inverse kinematics reaction is that the products are
strongly focused along the beam direction and, thus, can be easily collected 
for immediate use.
%%%

Proof-of-principle of the aforementioned approach is presented for the production of $^{67}$Cu 
via the reaction of a $^{70}$Zn beam at 15 MeV/nucleon with a hydrogen gas target.
%%%
This work demonstrates that important non-standard medical radioisotopes  %%% like $^{67}$Cu 
with high radionuclide purity can be produced, provided that low-energy and  
high-intensity primary beams are available.
%%%
Our method has some similarity with the isotope-harvesting approach from fragmentation facilities, 
in the sense that, in the latter, the fragmentation reactions also occur in inverse kinematics, 
albeit at high energies (above 100 MeV/nucleon) producing a very broad range of isotopes.
The inverse kinematics approach, however, takes place at low energy. 
%%%
By choosing the appropriate reaction channel,  
the radionuclide of interest  can be selectively produced with minimal radioimpurities
and implanted in an appropriate  catcher material for subsequent use 
(after minimal radiochemical processing if necessary).
In parallel, the forward-focused neutrons from the primary reaction can be sent to a secondary 
target for additional radioisotope production. %%% as in (\cite{Auditore-2017}).

The structure of the paper is as follows: in section 2, we present the experimental setup and 
the measurements; in section 3, we continue with the data analysis and the results. 
In section 4,  we discuss  improvements of the present method and further plans. 
%%%
Finally, in section 5,  we provide a summary and conclusions.

%%**********************************************************************************************
%%**********************************************************************************************
%%% Fig 1 Gas cell

\begin{figure}[h]   %%% htbp
\begin{center}  
\includegraphics[width=0.30\textwidth,keepaspectratio=true]{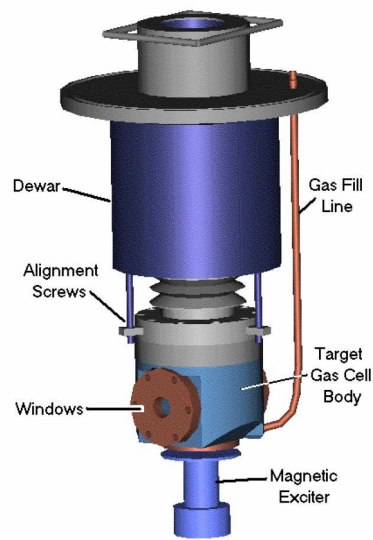}
\end{center}  
\caption{ (Color online)
%%%
The cryogenic gas target cell used in the present work (\cite{Brinkley-2003}). 
%%%
}
\label{figure01}
\end{figure}
%%**********************************************************************************************

%%**********************************************************************************************
%%**********************************************************************************************
%%% Fig 2 Schematic diagram of the setup

\begin{figure*}[h]   %%% htbp %%% CAUTION in the "*" for two column figure
\begin{center}  
\includegraphics[width=0.70\textwidth,keepaspectratio=true]{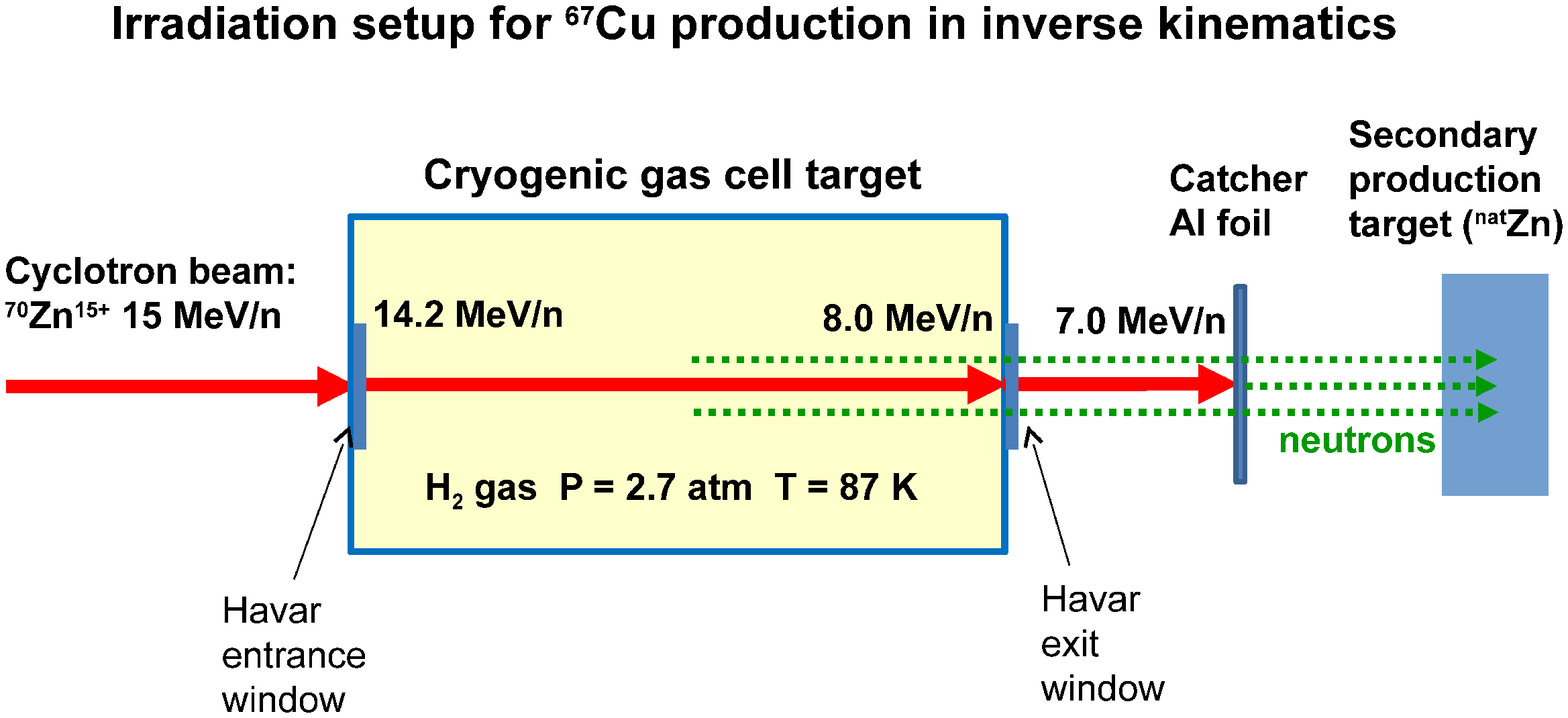}
\end{center}  
\caption{ (Color online)
%%%
Schematic diagram of the irradiation setup. 
A $^{70}Zn$ beam at 15 MeV/nucleon enters the gas cell and interacts with the hydrogen gas.
The heavy reaction products, including $^{67}$Cu,   after exiting the gas cell  are implanted 
in the Al catcher. 
%%%
The energies of the beam are listed as it passes through the entrance window, the gas and the 
exit window (thick red arrows). 
%%%
The dashed (green) arrows represent the neutrons produced via the interactions of the $^{70}Zn$ beam 
with  the hydrogen gas and the Al catcher.
%%% 
For details, see sections 2, 3.2 and 3.3.
%%%
}
\label{figure01}
\end{figure*}  %%% CAUTION in the "*" for two column figure
%%**********************************************************************************************

\section{Experimental Setup and Measurements}

The experimental work took place at the Cyclotron Institute of Texas A\&M University (TAMU).
%%%% (\cite{TAMU-2018}) %%% 2017-2018 Progress Report
A primary beam of $^{70}$Zn$^{15+}$ from the ECR source was accelerated 
by the K500 superconducting cyclotron to an energy of 15 MeV/nucleon and 
transported to the target chamber of the MARS recoil separator (\cite{MARS}).
%%%
The beam impinged on a cryogenic gas cell filled with H$_{2}$ gas 
held at a pressure of 2.7 atm  in contact with a liquid nitrogen reservoir (\cite{Brinkley-2003}).
%%%%
The cryogenic gas cell (figure 1) has a length of 10 cm with 4 $\mu$m Havar entrance and exit windows 
of 19.0 mm diameter.
%%%
%%The cell is operated at LN2 temperature to increase the gas density 
%%without need to increase the window thickness.
%%
The experimental setup is schematically shown in figure 2. 
An aluminum foil, mounted on an aluminum target frame (with a hole of 12.7 mm) 
placed after the hydrogen gas cell, was used  to collect the produced
 $^{67}$Cu nuclei from the reaction  of the $^{70}$Zn beam with the proton target.  
%%%
The irradiation lasted 6.5 h with a beam current of 0.31 pnA (particle nA) (2.0$\times$10$^{9}$ particles s$^{-1}$).
The current was periodically monitored and was nearly constant (within 15 \%).
%%%
The measurement of the current was performed by inserting a Faraday cup mounted on the 
same target ladder as the Al catcher frame. The measurement of the beam current at this location
(i.e. after the gas cell) was 8.0 nA of $^{70}$Zn (7.0 MeV/nucleon) at an average charge state of 26+.
This equilibrium charge state value was calculated with the parametrization 
of Leon et al. (\cite{Leon-1998}).
%%%
After 36.4 h from the end of the irradiation, the Al catcher foil was moved
in front of a high-purity germanium (HPGe) detector for 
off-line $\gamma$-ray analysis as described in the following sections.

%%%
%%%--------------------------------------------------------------------------------
%%% measured activity:

\section{Data analysis and results}

\subsection{ Off-line $\gamma$-ray analysis }

The radioactivity of the produced $^{67}$Cu and the other coproduced radioisotopes
was determined by off-line analysis of the  $\gamma$-ray spectra.
The foil was placed at a distance of d = 17.2(10) mm from the end cap of the detector.
Under this condition, the dead time of the counting system was around 2-3\%, thus avoiding
the pile-up effect. 
The energy resolution of the detector system was 2.5--4.0 keV FWHM.
%%%
%% The $\gamma$-ray spectra of the radionuclides were obtained using a pre-calibrated 
%% HPGe detector coupled to a PC based multichannel analyzer. 
%% The energy resolution of the detector system was 3.0 keV FWHM at 
%% the 1332.5 keV $\gamma$-ray peak of $^{67}$Co. 
%%%The energy and efficiency calibration of the HPGe detector system was
%%performed by using a standard $^{152}$Eu source. 

The energy calibration was performed using known $\gamma$-rays obtained in the spectra.
The absolute efficiencies were obtained using photopeak efficiency predictions generated from a Geant4 
(\cite{Agostinelli-2003}) simulation, considering the source and the detector geometry.
The spectrum of the room background was measured for 67.3 h.
%%
%% Spectroscopy software ??? was used for the analysis. 
%%

The radioactivity levels of the isotopes were determined by the quantification of the
photopeaks of the $\gamma$-rays
%%
%% $^{67}$Cu (184.6 keV),
%% $^{65}$Zn (1115.5 keV, T$_{1/2}$=244d), 
%% $^{69m}$Zn (438.6 keV, T$_{1/2}$=13.8h), 
%% and $^{64}$Cu (1345.8 keV,T$_{1/2}$=12.7h),
%%
taking into account the branching ratios and the absolute efficiencies of the detector.  
A detailed description of the  $\gamma$-ray analysis of  all the observed $\gamma$-ray 
peaks will be presented in \cite{Rodrigues-2019}.

%%% Equation:
%% using the following equation:	 Activity in Bq = (N obs ) ⋅ D(RT) /( ε ⋅ I γ ⋅ LT) 
%% where N obs  = number of detected photo-peak counts of the γ-ray energy, 
%% D(RT) = decay factor to correct decay during counting time = λ/(1–e– λRT ), 
%% I γ  = the branching intensity or the abundance of the γ-ray of interest, 
%% ε = experimental efficiency of the HPGe detector for the γ-ray energy considered, 
%% LT = live counting time, RT = real time, λ = decay constant (=0.693/T 1/2 ) 
%% of the radionuclide of interest with half-life T 1/2 .
%%%

%%**********************************************************************************************
%%**********************************************************************************************
%%% Fig 3   Gamma spectrum

\begin{figure*}[h]   %%% htbp  %%% CAUTION in the "*" for two column figure
\begin{center}  
\includegraphics[width=1.05\textwidth,keepaspectratio=true]{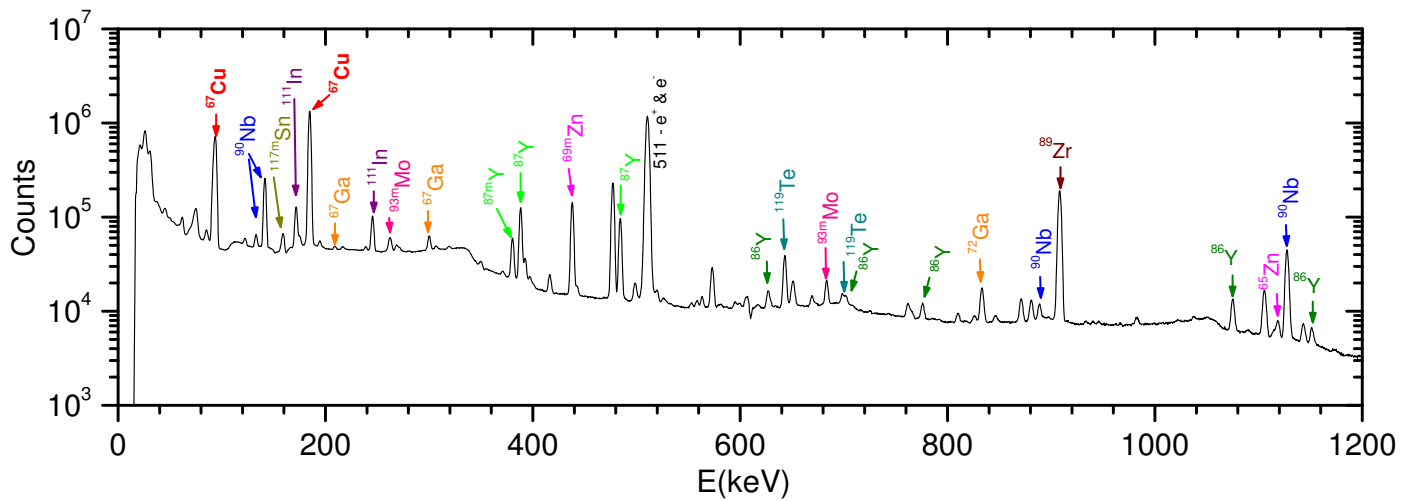}     %% 1.00 is OK
\end{center}  
\caption{ (Color online)
%%%
Typical background-subtracted gamma-ray spectrum of the Al catcher foil 
following the interaction of a $^{70}$Zn (15 MeV/nucleon) beam with the hydrogen gas cell 
after 36.4 h from the end of the irradiation.
%%%
The peaks are labelled with the corresponding radionuclides (see text).
%%%
We specifically note the two peaks (at 92 keV and 185 keV) due to $^{67}$Cu. 
%%%
and the peak (at 439 keV) due to $^{69m}$Zn (T$_{1/2}$ =13.8 h), which is 
the main radioimpurity coming from the production reaction $^{70}$Zn+p.
%%%
}
\label{figure03}
\end{figure*}    %%% CAUTION in the "*" for two column figure

%%**********************************************************************************************
%%**********************************************************************************************
%%% SPECTRUM:

In figure 3, we present the background-subtracted $\gamma$-ray spectrum  obtained 
during an accumulation period of 68.0 h, starting 36.4 h  after the end of the irradiation. %%% (GEORGE, verify).
%%%
We note that the peaks at 92 keV and 185 keV are characteristic of $^{67}$Cu decay.
The peak at 439 keV is due to the main radioimpurity of $^{69m}$Zn (T$_{1/2}$ =13.8 h)
from the $^{70}$Zn+p reaction.   %%%%  (see below).
%%%
We observe a small contribution, around 3\% of the $^{67}$Cu production, from $\gamma$-ray at 300.2 keV 
characteristic of the presence and decay of $^{67}$Ga (T$_{1/2}$=78 h) with a branching ratio of 16.7\%. 
%%%
We note that this radioimpurity, decaying to the same levels of $^{67}$Zn as $^{67}$Cu,
is intensely coproduced in the main production route via $^{68}$Zn(p,2p)$^{67}$Cu with 
high energy protons (\cite{Smith-2012}). This $\gamma$-ray for $^{67}$Cu has a branching ratio of only 0.8\%.
%%%
Similarly, the peak at 1115.5 keV, characteristic of $^{65}$Zn (T$_{1/2}$ =244 d), is significantly 
suppressed (0.2\% of the  $^{67}$Cu production).
This radionuclide is notoriously produced in the high-energy $^{67}$Cu production methods and requires 
radiochemical separation.

In the spectrum of figure 3, we observe peaks from a series of radionuclides that are mainly due
to nuclear reactions on the Havar windows and the Al catcher foil. More specifically,
the radionuclides $^{86}$Y, $^{87}$Y, $^{89}$Zr, $^{90}$Nb and $^{93m}$Mo are fusion-evaporation 
products of the reaction $^{70}$Zn+$^{27}$Al 
(with the beam of $^{70}$Zn entering the Al catcher at 7.0 MeV/nucleon, see section 3.2).
%%%
Furthermore, the heavier radionuclides $^{111}$In,  $^{117m}$Sn and  $^{119}$Te are residues  of the
reaction of the $^{70}$Zn beam with the constituents of the Havar alloy (42\% Co) of the gas-cell windows.
%%%
We also identified the presence of $^{7}$Be (T$_{1/2}$=53 d, E$_{\gamma}$=477.6 keV) possibly coming 
from the activation of the  small Be or C content of the Havar alloy (0.02--0.08\%, 1.6\%, respectively). 
%%Finally, the radionuclide  $^{24}$Na (T$_{1/2}$ =15 h) may result from nucleon stripping 
%%reactions on the Al catcher.  %%% gamma's: 1367, 2754 KeV. 
A complete analysis of the origin and the activity of these radionuclides will be presented in 
(\cite{Rodrigues-2019}).
%%%
As we discuss in section 3.2, the radioimpurities coming from the Havar windows and the 
Al catcher can be easily reduced or completely eliminated by proper tuning of the energy loss of the
primary beam in the gas cell, so they do not present a potential problem in the proposed method
of radionuclide production.
%%%-------------------------------

From the above analysis, the activity of the produced $^{67}$Cu at the end of the 
6.5 h of irradiation was 1.6(5) kBq with the beam current of 0.31 pnA. 
This leads to a specific activity of 0.8(3) kBq/h/pnA  (or equivalently, 0.8(3) MBq/h/p$\mu$A).

Along with the activity of $^{67}$Cu, we report the measured activity of 0.094 kBq of the 
$^{69m}$Zn (T$_{1/2}$ =13.8 h) at the end of the irradiation (this corresponds to a specific activity
of 0.045 kBq/h/pnA).
%%%
%We notice that this is higher that the specific activity of $^{67}$Cu; we, thus, infer that an appropriate 
%cooling period (2--3 days) should be necessary for this activity to reach an acceptably low level.
%%
This is a rather low activity that can be further reduced by an appropriate cooling period 
(of, e.g., 2 days).

\subsection{Comparison of measured $^{67}$Cu activity with estimates}

In this section, we estimate the activity of the radionuclide $^{67}$Cu produced in the
inverse kinematics reaction $^{70}$Zn (15 MeV/nucleon) + p.
%%%
The cross sections for the direct kinematics reaction 
$^{70}$Zn(p,$\alpha$)$^{67}$Cu have been reported in \cite{Kastleiner-1999}.
%%%
We take into account that the 10 cm long gas target cell was filled with H$_{2}$ 
gas at a pressure of 2.7 atm  and was at a temperature of 87 K via the thermal contact 
with the LN$_{2}$ Dewar. Under these conditions, the H$_{2}$ gas target thickness was 7.5 mg/cm$^2$.
%%%
Using the code SRIM (\cite{Ziegler-2013}), we calculated the energy losses of the beam in 
the entrance window, the gas and the exit window of the gas cell.
%%%
The $^{70}$Zn beam hit the entrance window at 15 MeV/nucleon, exited it at 
14.2 MeV/nucleon, traversed the gas and reached the exit window at 
8.0 MeV/nucleon and, finally exited it at 7.0 MeV/nucleon (figure 2).

Using the data  of Kastleiner et al. (\cite{Kastleiner-1999}),
we estimated the integral of the excitation function in the region 14.2--8.0 MeV,
resulting in an average value for the cross section of 7.5 mb  for $^{67}$Cu.
%%%
Using this cross section value and the H$_{2}$ gas target thickness,  
we obtained the production rate of $^{67}$Cu. 
%%%
Subsequently,   we obtained the produced activity, assuming 1h of irradiation with a beam 
intensity of 1 pnA. This activity is calculated to be 1.8 kBq/h/pnA 
(or equivalently, 1.8 MBq/h/p$\mu$A).
%%%
We notice that the measured activity of 0.8(3) kBq/h/pnA is about 2.2 times lower 
that the expected activity, as obtained with the procedure described above.
%%%------------------------

%%Before examining the origin of this discrepancy for $^{67}$Cu, we wish to 
%%present a consistency test for the ratio of 1.3 of the measured specific activities of 
%%$^{69m}$Zn (1.08 kBq/h/pnA ) and $^{67}$Cu (0.83 kBq/h/pnA), as we mentioned previously.
%%%
%%To obtain this ratio theoretically, we need the production cross sections of the isotopes
%%in the relevant energy region. For $^{67}$Cu, we already gave the value above. For $^{69m}$Zn,
%%we used TALYS to obtain the production cross sections at energies from 15 MeV down to 11 MeV
%%(where the cross section is found to be practically zero). In the energy interval of 14.2--11.0 MeV,
%%we obtained a mean value of about 3.0 mb. We note that this energy interval exploits only 
%%3.8 mg/cm$^2$ of the full 5.6 mg/cm$^2$ of the target thickness (or, equivalently, the length of the gas cell).
%%Taking into account these estimates and the half-lives of the isotopes,  we calculated  a ratio
%%for their ativities of 1.2, which is in very good agreement with the measured ratio.   

%%%------------------------

In this section, we discuss several possible sources for the discrepancy between the experimental 
and the theoretical activity of $^{67}$Cu:
%%%
a) the spread of the recoiling $^{67}$Cu nuclei that may result in an  incomplete collection
  on the Al catcher foil. As previously mentioned, the Al foil was mounted on an Al frame with a 12.7 mm hole. 
  The $^{67}$Cu activity of the target frame was  measured to be 0.12 kBq which is 7.5\% of the main activity
  collected on the Al catcher foil;
%%%
b) the use of an unsuppressed Faraday cup to measure the beam current. As it is well known,
  the sputtering of electrons from the side of the Faraday cup hit by the heavy-ion beam
  results in an increase of the measured positive current leading to a lowering of the
  obtained production rate and, thus, the specific activity. The cup should be biased to a positive
  voltage (in the range 200--400V). Such electron suppression of the Faraday cup was not performed 
  in the present experiment.
  %%%
  This effect may account for about 20--30\% of the discrepancy, but it should be quantified for the 
  present heavy-ion setup  (as performed in \cite{Carzaniga-2017}, figure 4, for proton beams);
%%%
c) the reduction of the pressure of the H$_{2}$ gas due to local  heating along the path 
  of the primary beam. As reported in a previous study (\cite{Brinkley-2003}), this effect has been 
  observed  in the behavior of this gas cell during its use for radioactive beam production
  with the MARS recoil separator. The rate of radioactive beams has been observed to drop 
  when intense primary beams were used, but no quantitative estimates have been reported.
%%%
A related quantitative account of this effect is reported by Amadio et al. 
(\cite{Amadio-2008}) for the production of a low-energy radioactive beam of $^{7}$Be
in a cryogenic hydrogen gas cell. In that setup, circulation of the hydrogen gas cooled 
at LN$_{2}$ temperature indicated that the effect of the local pressure reduction may 
amount to 60--70\%  at high primary beam currents [figure 4 of (\cite{Amadio-2008})].
%%%
Similar measurements are necessary for the present gas cell. We note that a magnetic stirring
system was operated in this cell, but its effect was rather inadequate.

Finally, we mention that, by using the cross section data of  (\cite{Kastleiner-1999}),
the production cross section of $^{67}$Cu drops to nearly zero at about 7 MeV. 
Thus, in the energy range 14.2--8.0 MeV/nucleon of the $^{70}$Zn beam in the hydrogen gas,
almost the full thickness of the target was exploited to produce $^{67}$Cu.

\subsection{Use of the neutrons from the primary reaction for secondary radioisotope production}

The interaction of $^{70}$Zn at 15 MeV/nucleon with the proton target produces 
about 1.6 neutrons per reaction, as calculated with the code TALYS (\cite{Koning-2012,Duchemin-2015}). 
%%%(GEORGE: give indicative energy and ang. dist. info).
%%%
Along with the heavy reaction products, these neutrons are also kinematically focused in the forward direction
and can be exploited for further radioisotope production by simply letting them interact with a  secondary target, 
e.g., a $^{68}$Zn or $^{nat}$Zn target for additional production of $^{67}$Cu. 
%%%
In this work, we performed such a test by placing a block of twenty 25.4x25.4 mm$^{2}$ foils of $^{nat}$Zn with 
1 mm thickness behind the Al catcher (figure 2).
%%%

Because this production is from the secondary neutrons, the yield is expected to be about 
2 orders of magnitude lower than the main production channel. 
However, given  the long mean free path of the neutrons (with a typical value of several cm), 
production of different isotopes is conceivable with these neutrons impinging on different targets. 
%%%
This production scheme shares some similarity with the one recently published in (\cite{Auditore-2017}) 
employing secondary neutrons from the target of a standard $^{18}$F radioisotope-producing setup.
%%%
The advantage of our approach of inverse kinematics is the strong focusing of the 
secondary neutrons that can be directed to a target stack of much smaller size than the one used 
in (\cite{Auditore-2017}),  which essentially encloses the target.
%%%
The results of our neutron-production test are promising. The analysis work is currently in progress 
and will be presented in \cite{Rodrigues-2019}.  
%%%

%%%***************************************************************************
%%----------------------------------------------------------------------------

\section{Discussion and plans}

The present preliminary study that we performed at the Cyclotron Institute of TAMU 
confirms that important medical radionuclides, such as $^{67}$Cu, 
can be effectively produced using inverse kinematics.  
The main advantages of the present novel  approach along with necessary developments
and relevant implementations  are discussed below.

First, the produced radionuclides are strongly focused along the beam direction and,
thus, can be easily collected. %%%  for analysis and immediate use.
%%%
In this respect, with the appropriate choice of the reaction channel(s) and the subsequent cooling
time of the products, it is possible to minimize the production  of radioimpurities
coming from the main reaction. Moreover, it is possible to minimize the radioimpurities
resulting from the primary  beam interacting with the Havar windows of the gas target 
and the catcher material. 
%%%
Of course, we cannot avoid reactions at the entrance window, where the beam enters with
the full energy of 15 MeV/nucleon and induces reactions on the foil.
%%% 
However, the products coming from peripheral or semiperipheral (deep-inelastic) collisions 
on the isotopes of Havar (Co, Cr, Ni, W, etc.) have rather  wide angular distributions
[e.g., (\cite{Fountas-2014}) and (\cite{Papageorgiou-2018}) figure 4]  
and are expected to mostly miss the catcher (depending on its diameter). 
%%%
On the other hand, the products of complete or nearly complete fusion are forward focused, 
but are heavier and slower than the beam and, thus, may mostly stop in the gas.

Regarding the exit window, it is possible to adjust the gas cell parameters
(pressure, temperature and length)  so that the primary beam reaches this window  
at low energy, i.e. near or below the Coulomb barrier  of the relevant reactions
(e.g., 4.0 MeV/nucleon).   Consequently,  nuclear reactions can be  suppressed or 
fully eliminated on that window.
%%%
Of course, under these conditions, the low energy primary beam exiting 
the gas cell  will not induce reactions in the catcher material.
For example, referring to the $^{70}$Zn+$^{27}$Al reaction, the Coulomb barrier corresponds 
to a projectile energy  of 3.5 MeV/nucleon.
%%%
We realize that detailed simulations and further experimental tests are necessary to achieve optimum 
conditions for the experimental setup and production procedure.
%%%
Under properly fine-tuned conditions,  water or other materials (salt, sugar, etc.) can be used 
to collect the radioisotopes in a convenient chemical form, so that,  post radiolabelling, 
they may be used for tests on animals.

%%%-------------------------------------------------------------------
%%% Neutrons, discussion:

As we briefly mentioned, secondary neutrons from the primary reaction can be used to irradiate other targets
for further radioisotope production of the same or different type (e.g. Cu, Sc, etc.). 
However, in this case, radiochemical methods are needed to separate the medical radionuclides,
as in the traditional production schemes (\cite{Smith-2012}).

%%%-------------------------------------------------------------------
%%% Cost reduction : need of a HI accelerator:
From a financial point of view,  material costs may be considerably reduced,  since the heavy 
(and usually rare) element is used as the projectile 
(for instance, $^{70}$Zn has 0.6\%  natural abundance).
%%%
Also, the radiochemical processing is substantially minimized or, desirably, completely eliminated 
because the radioisotope of interest is collected and essentially used directly after production
(and appropriate cooling).
%%%
However, the primary requirement  of  our approach is the use of a heavy-ion accelerator 
(e.g., cyclotron or LINAC) that can deliver high-intensity heavy-ion beams in the energy range  10--20 MeV/nucleon. 
%%%
Fortunately, such accelerators are available at a number of facilities worldwide and, with appropriate planning, 
a fraction of their beam time may be devoted to non-standard radionuclide production following 
our inverse-kinematics approach.
%%%-------------------------------------------------------------------

According to our estimates, with a primary beam of 1 particle $\mu$A,  we can reach activities of 1.8 MBq/h 
and, thus, obtain milliCurie quantities of $^{67}$Cu within 24 h of irradiation.
%%%
The p$\mu$A heavy-ion beam intensities are achievable with the current ion-source and accelerator technology.

We note that the use of the cryogenic gas cell has the additional advantage that its cooled windows 
can withstand the necessary high beam currents.
%%%
Appropriate circulation of the cooled hydrogen gas will be necessary to mitigate the effect of 
the local denstity reduction.
%%%
%%% Lower temperature operation:
%%%
For safe operation of the gas cell under intense beam irradiation, even with thinner windows, 
we may consider lowering the pressure to 1.0--1.5 atm, increasing the length, and, furthermore, 
lowering the temperature below LN$_{2}$ temperatures with the use of modern cryocoolers 
[e.g., a Gifford-McMahon refrigerator (\cite{cryocoolers-2009})].

%%% LIQUID H2 :
%%%
Taking advantage of the development of thin liquids, it is conceivable to substitute the hydrogen 
gas cell by a liquid H$_{2}$ cell with thin windows. 
%%%
The development of a liquid H$_{2}$/D$_{2}$ target with typical thickness of a few mm is reported in 
(\cite{Jaeckle-1994}).
%%%
More recently, a liquid H$_{2}$ target for fragmentation reactions has been reported in (\cite{Ryuto-2005}).
This target was operated at temperatures about 20 K, achieved with a Gifford-McMahon refrigerator.
The cell had a length of 30 mm and a density of 200 mg/cm$^2$.
%%%
We propose the implementation of a similar system for the production of medical isotopes, 
but for this  purpose the length has to be only 1-2 mm to achieve the required thickness 
of about 8--10 mg/cm$^2$.

%%%-------------------------------------------------------------------
%%% TAMU CI:
In the case of the Cyclotron Institute of TAMU, the facility houses two cyclotrons: 
%%%
a) the K150 cyclotron which, in principle, can produce high-intensity heavy-ion beams  (up to around Kr)
with energies 
up to around 12--15 MeV/nucleon, suitable for the production of relatively large activities of radioisotopes and,
%%%
b) the K500 cyclotron, employed  in the present experiment,  which can produce lower currents of heavy-ion beams
(upto $^{238}$U) and in a broader energy range (up to 20--40 MeV/nucleon depending on the isotope). 
%%%
We anticipate that both the K150 and the K500 cyclotrons may  be successfully used for the development and 
production of a  variety of non-standard radioisotopes at activities appropriate for medical studies on
small animals.
%%%
For this purpose, however, beams of 30--100 pnA have to be developed, which should be possible  with a modest 
investment at the existing ion-source and accelerator infrastructures.

%%% CLOSING:
%%%
To summarize,  the aforementioned considerations  for the development of a viable route of medical radioisotope
production in inverse kinematics are based, first,  on the successful results of the present study and, second, 
on the current experience and developments on ion-source and accelerator technologies worldwide.
%%%
We think that a fruitful application of our proposed method is acheivable with timely planning and 
allocation of relatively modest resources. 
%%%
Along with the other production approaches, the proposed route may  contribute to a broad and 
diversified program of production and use of non-standard medical radionuclides.

%%%***************************************************************************
\section{Conclusions}

%%% From the abstract:

An innovative  method for the production of important medical radioisotopes was presented  
in this article. 
%%%
The approach is based on realizing the nuclear reaction in inverse kinematics by 
sending a heavy-ion beam of appropriate energy on a light target (e.g., H, d, He) and 
collecting the isotope of interest on an appropriate catcher after the target.
%%%
In this work, as a proof-of-principle, we studied the production of the 
radionuclide $^{67}$Cu (T$_{1/2}$=62 h) 
via the reaction of a beam of  15 MeV/nucleon $^{70}$Zn with a cryogenic hydrogen gas target.
%%% 
The $^{67}$Cu radionuclide (along with other coproduced isotopes) was collected 
after the gas target on an Al catcher foil and the radioactivity was measured by off-line 
$\gamma$-ray analysis. 
%%%
After the end of the irradiation, the main radioimpurity  in the Al catcher coming from 
the $^{70}$Zn+p reaction was  $^{69m}$Zn (T$_{1/2}$=13.8 h),  which can be suppressed 
by cooling for a period of 2--3 days. 
Other identified radioimpurities are understood to come from the interaction of the beam with 
the window material and the catcher and can be eliminated by careful tuning of the parameters 
of the setup.
%%%
The present successful test and the ensuing considerations indicate the possibility of 
producing important non-standard  radionuclides %%% like $^{67}$Cu, 
of high radionuclide purity with the approach of inverse kinematics.
%%%
In parallel to the main production scheme, secondary neutrons from the primary reaction
were used to irradiate a secondary target of Zn for further radioisotope production 
with promising results.
%%%
The main requirement necessary to achieve production of activities appropriate for preclinical
studies is the availability  of high-intensity (particle $\mu$A) heavy-ion primary beams.

%%%***************************************************************************
\section{Acknowledgements}

We are grateful to the support staff of the Cyclotron Institute for providing 
the primary beam. 
%%%
%% We are also thankful to .... for enlightening comments
%% and suggestions on this work. 
%%%
%%% Finally, we wish to acknowledge motivating discussions with....and other
%%% members of the RISP facility. 
%%%
We are thankful to Dr. H.T. Chifotides for inspiring  discussions and 
suggestions on this collaborative work and for critical reading and editing of 
the manuscript.
%%%%
%%% Financial support for this work was provided, in part, 
%%% by the US Department of Energy  Grant No. DEFG02-93ER40773 (???),
%%% the Welch Foundation under Grant No. A-1397 (???)
%%%
Financial support for this work was provided,
in part by NNSA under grant no de-na0003841 (CENTAUR),
and by the National and Kapodistrian University of Athens
under the ELKE Research Account. %%%  No 70/4/11395.
MY and PZF acknowledge the support of Houston Methodist 
Research Institute (HMRI).
%%%
%%%***************************************************************************
%%-----------------------------------------------------------------------------
%% The Appendices part is started with the command \appendix;
%% appendix sections are then done as normal sections
%% \appendix

%% \section{}
%% \label{}
%%------------------------------------------------------------------------------
%% If you have bibdatabase file and want bibtex to generate the
%% bibitems, please use
%%
%%  \bibliographystyle{elsarticle-harv} 
%%  \bibliography{<your bibdatabase>}

%% else use the following coding to input the bibitems directly in the
%% TeX file.

%%%***************************************************************************

%%----------------------------------------------------------------------------

\end{document}